\documentclass[a4paper,11pt]{article}
\usepackage{pos}
\newcommand{\noop}[1]{}

\title{The intriguing outcomes of double degenerate mergers}

\author*[a]{Lilia Ferrario}

\affiliation[a]{Mathematical Sciences Institute, The Australian National University,\\
  Hanna Neumann Building 145, ACT2601, Canberra, Australia}

\emailAdd{Lilia.Ferrario@anu.edu.au}

\abstract{This review talk explores the diverse outcomes of white dwarf mergers, emphasising that not all double degenerate mergers result in supernovae. Possible outcomes include the formation of a more massive WD, partial explosions that leave behind unusual remnants, and merger-induced collapse leading to the formation of rapidly spinning neutron stars or magnetars. Additionally, merging WDs have been proposed as potential progenitors for long $\gamma$-ray bursts and Fast Radio Bursts.
}

\FullConference{Frontier Research in Astrophysics – IV (FRAPWS2024)\\
9-14 September 2024\\
Mondello, Palermo, Italy\\}

\usepackage{fancyhdr}
\usepackage{graphicx}
\usepackage{placeins}
\usepackage{pdfpages}
\usepackage{subcaption}
\usepackage{caption}
\usepackage{color}
\usepackage{multirow}

\newcommand{\msun}{M$_{\odot}$}

\begin{document}
\maketitle

\section{Introduction}

Merging stars in binaries are common events in the universe, but the outcomes of these mergers remain under debate due to the many uncertainties arising from the wide range of possible initial conditions (e.g., masses, separations, stellar rotations, metallicity) and aspects of binary evolution that are often not yet fully understood, such as the common envelope phase.

Only about 25\% of white dwarfs (WDs) in the local sample are found in binary systems (\cite{Ferrario2012}), compared to an approximately 50\% incidence of binarity observed among intermediate-mass stars on the main sequence. While \cite{Ferrario2012} proposed that this discrepancy might be due to many WDs being obscured by the glare of their more luminous companions and thus missing from the WD census, \cite{Toonen2017} showed that the low percentage of WDs in binaries is primarily caused by post-main-sequence merging events.

Merging can occur at different stages of the evolution of a binary, e.g., during one of the common envelope phases of binary evolution, or as two compact stars are drawn together by gravitational wave (GW) emission. Potential outcomes include:
\begin{enumerate}
    \item the formation of a more massive WD (\cite{Dunlap2015, KawkaNonexplosive2023, Ferrario1997_J0317});
    \item supernovae of type Ia (SNe\,Ia), routinely used to probe the geometry and the accelerating expansion of the universe (\cite{Schmidt1998,Riess1998,Perlmutter1999});
    \item an explosion that leaves behind a peculiar WD which could be a fragment of the exploding WD  (\cite{Vennes2017,Raddi2019MNRAS}) or its runaway companion WD (\cite{shenThreeHypervelocityWhite2018}, \cite{El-Badry2023_D6stars});
    \item a merger-induced collapse leading to the formation of a rapidly spinning neutron star \cite{Ruiter2019}.
    \item the formation of long $\gamma$-ray bursts (\cite{Tout2011}) and Fast Radio Bursts (\cite{lorimerBrightMillisecondRadio2007}).
\end{enumerate}

In the following sections, I will review some of the different channels that lead to some of the most spectacular and powerful astrophysical events and to the formation of exotic remnants. The key characteristics of a class of WDs that result from the merger of two WDs are outlined in Section \ref{properties}. In Section \ref{SN_Progenitors} it is discussed how double-degenerate mergers represent a significant pathway for SNe\,Ia explosions. In Section \ref{long_bursts} it is proposed that the merging of two WDs can lead to the generation of rapidly spinning magnetars which could be associated with the phenomena that surround the formation of long $\gamma$-ray bursts (LGRBs). In Section \ref{formation_FRB} it is suggested that the merging of double WD systems could be associated to the formation of Fast Radio Bursts. The main findings are summarised in Section \ref{Conclusions}.

\section{Merged White Dwarfs}\label{properties}

WDs that are the outcome of double WD mergers are expected to exhibit several distinct characteristics, namely:
\begin{enumerate}
    \item \textbf{Rapid Rotation:} Generally, WDs rotate slowly, with typical spin velocities of $\lesssim 40$\,km/s. However, those formed through double degenerate (DD) mergers are expected to rotate significantly faster, thus distinguishing themselves from the WD population that originates from single star evolution.
    \item \textbf{Magnetic Fields:} Strong magnetic fields can arise from a dynamo mechanism triggered by differential rotation during the merger process. It is thus expected that a larger incidence of magnetism should be observed among WDs that originate from DD mergers.
    \item \textbf{Masses and Peculiar Composition:} Post-merger WDs may have masses larger than the average mass of WDs ($\sim 0.6$\,\msun) and may display peculiar atmospheric chemical compositions due to nuclear burning and the loss of their envelope during the merging event.
\end{enumerate}

\subsection{Carbon-enriched DQ WDs}\label{DQ_WD}

\begin{figure}
    \centering
    \includegraphics[width=0.75\linewidth]{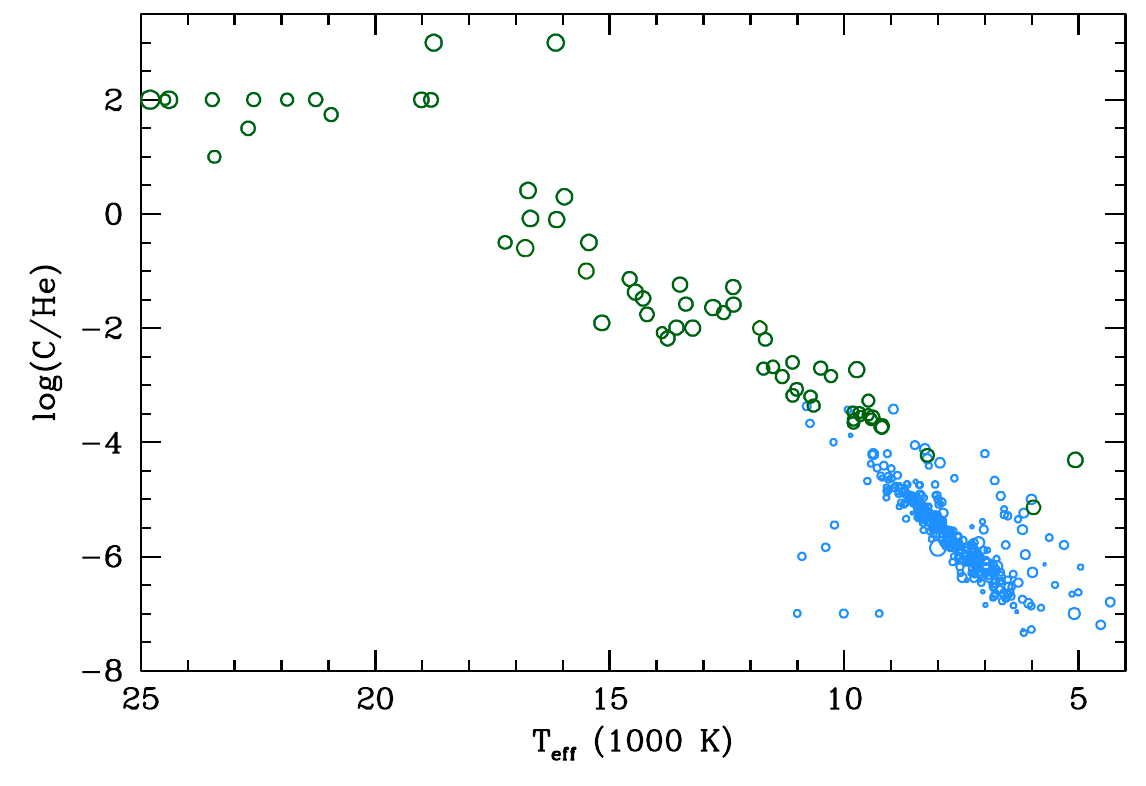}
    \caption{Carbon abundance against temperature of DQ WDs. The C-enriched DQs are plotted in green and the ordinary DQs in blue. The size of the circles is proportional to the mass of the WD (from \cite{KawkaNonexplosive2023}).}
    \label{C_vs_Temp}
\end{figure}

DQ WDs are a subclass of WDs whose spectra are dominated by atomic and molecular carbon. C-enriched DQ WDs not only exhibit spectral features similar to those of standard DQ WDs but also possess additional characteristics suggesting a completely different origin. In particular, they display the hallmarks of mergers, as outlined in Section \ref{properties}, along with kinematic properties typically associated with old Galactic populations. The characteristics and origin of these WDs have been thoroughly examined by e.g., \cite{KawkaNonexplosive2023, Coutu2019, Blouin2019, Koester2019}, and have revealed that they are denoted by relatively high temperatures ($10,000-25,000$,K), significant carbon enrichment, high masses ($M \ge 0.8$,M$_\odot$), and high incidence of magnetism ($\sim$ 70\% of the members of this class are magnetic). 

\begin{figure}
    \centering
    \includegraphics[width=0.5\linewidth]{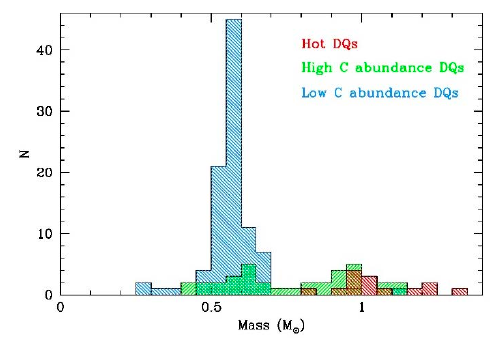}
    \caption{Mass distribution of DQ WDs. The ordinary DQs are in blue. The C-enriched warm DQs are in green and the hot ones in red (from \cite{Kawka2019_ASA}).}
    \label{DQ_Mass}
\end{figure}

\begin{figure}
    \centering
    \includegraphics[viewport=45 320 570 700,clip,width=0.75\columnwidth]{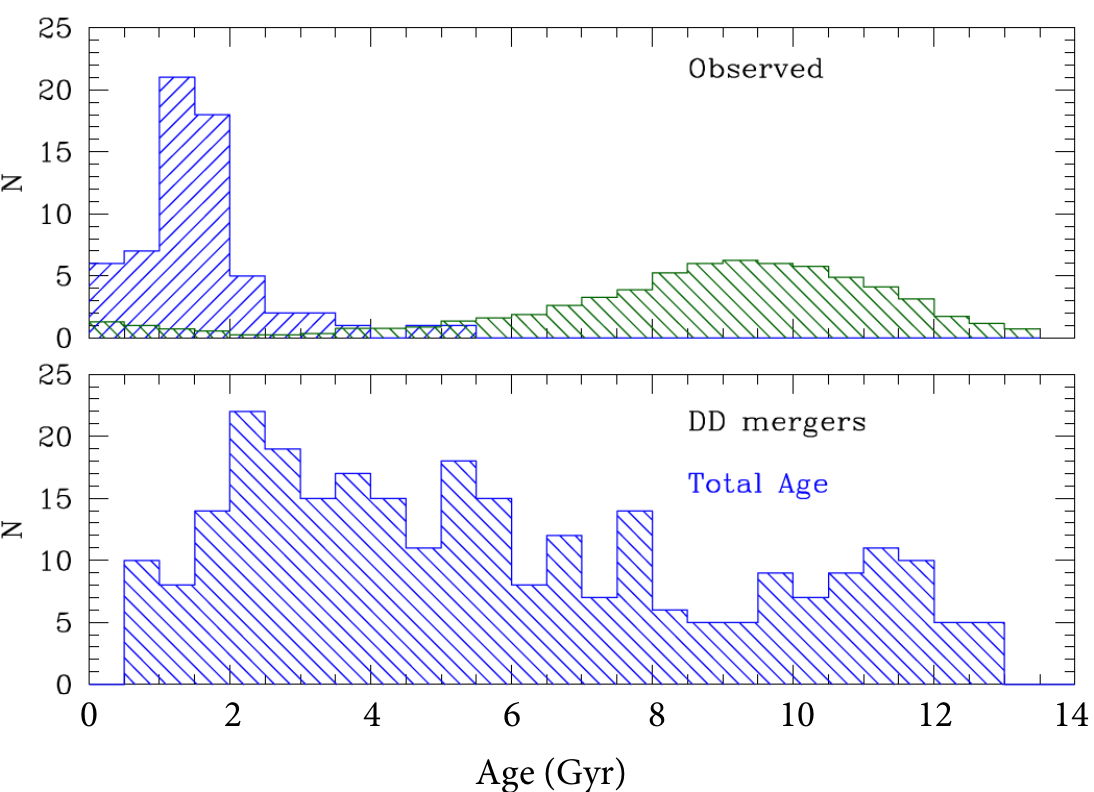}
    \caption{Top panel: Distributions of the measured cooling ages (blue) and of (real) ages determined by the eccentricity of their Galactic orbits (green). Lower panel: simulated age distributions for the DD mergers.The simulated data apply to the Gaia magnitude $G < 20$ synthetic sample (from \cite{KawkaNonexplosive2023}).}
    \label{fig_age_obs}
\end{figure}

Thus, these C-enriched WDs should not be confused with the ordinary cool DQ WDs, where helium convection brings carbon to the surface. Figure \ref{C_vs_Temp} highlights significant differences in carbon abundance, temperature, and mass between the two types of DQ WDs. Additionally, the mass distribution in Figure \ref{DQ_Mass} highlights that the hot and warm C-enriched DQs are much more massive than the ordinary DQs whose average mass is comparable to that of typical WDs arising from single star evolution.
C-enriched DQs appear young based on their temperatures and cooling ages, as illustrated by the blue histogram in the top panel of Figure \ref{fig_age_obs}, which shows their ages under the assumption of single-star evolution. However, their kinematics and spatial distributions suggest much greater ages, typical of very old populations such as the Galactic thick disc or halo, as shown by the green histogram in the same figure. This apparent contradiction can be resolved if these WDs formed through the merger of two WDs, producing remnants that retain the kinematic characteristics of the original double WD system while presenting as young WDs.

\begin{figure}
    \centering
    \includegraphics[width=0.75\linewidth]{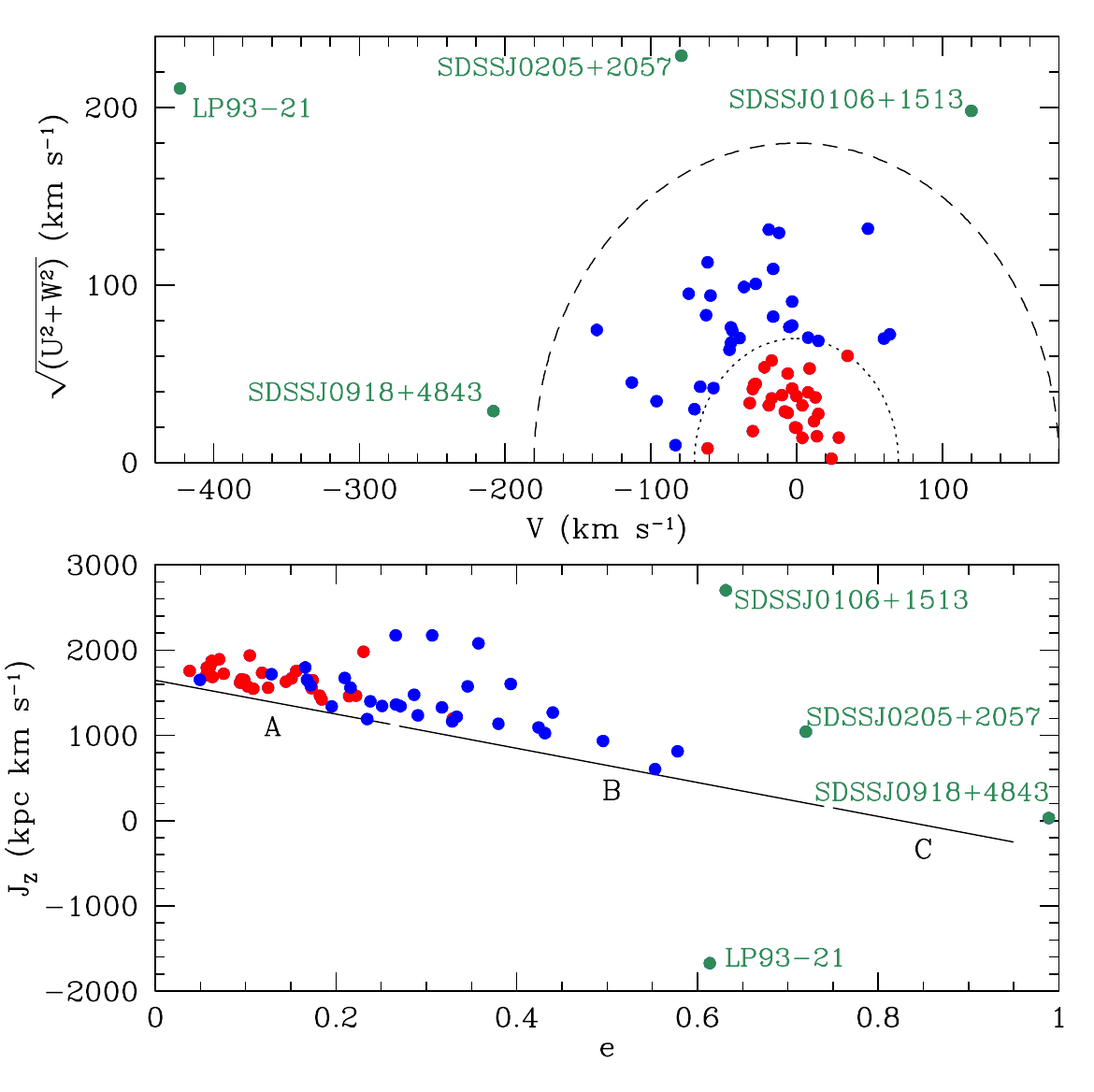}
    \caption{Top panel: Toomre diagram showing $\sqrt{U^2+W^2}$ versus $V$ of the C- enriched DQs. The dotted and short-dashed curves correspond to $\sqrt{U^2+W^2}$\,km\,s$^{-1}$ at 70, and 180\,km\,s$^{-1}$ respectively, and with respect to the local standard of rest. Bottom panel: Regions A , B , and C denote, respectively, thin disc, thick disc, and halo (from \cite{KawkaNonexplosive2023}).}
    \label{Toomre_DQ}
\end{figure}

The work of \cite{KawkaNonexplosive2023} shows that about half of the C-enriched DQ sample exhibit kinematic properties consistent with the Galactic thick disc or halo. This proportion is likely underestimated due to the assumption of radial velocities of $0$\,km\,s$^{-1}$ for a quarter of the C-enriched sample. The top panel of Figure \ref{Toomre_DQ} illustrates the velocity $\sqrt{U^2+W^2}$ versus $V$ (Toomre diagram), where $U$, $V$, and $W$ are the Galactic space velocity components. Stars can be broadly categorised based on their total velocity $v_t = (U^2 + V^2 + W^2)^{1/2}$. Those with $v_t\lesssim 70$\,km\,s$^{-1}$ (inner dotted circle) typically belong to the thin disc, while $70 \lesssim v_t \lesssim 180$\,km\,s$^{-1}$ indicate membership to the thick disc (outer dashed circle). Stars with $v_t \gtrsim 180$\,km\,s$^{-1}$ are generally considered halo objects.

\begin{figure}
    \centering
    \includegraphics[viewport=15 400 565 690,clip,width=0.75\columnwidth]{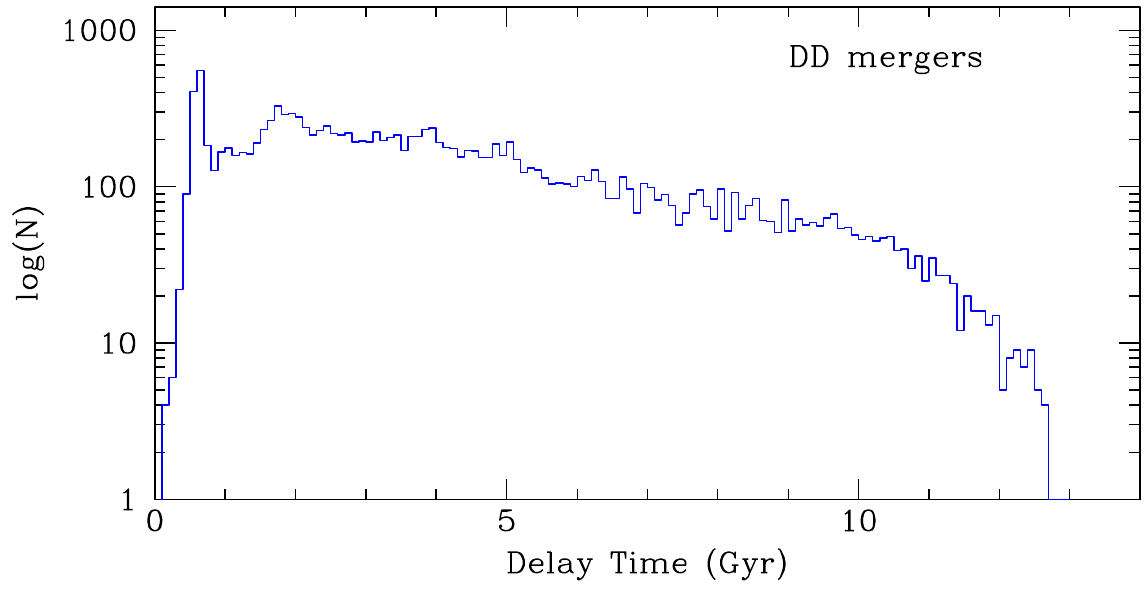}
    \caption{Distribution of delay times of the DD mergers within 1\,kpc from Earth and for one starburst only (from  \cite{KawkaNonexplosive2023}).}
    \label{delay_times}
\end{figure}

In the $J_z$ ($z$-component of the angular momentum) versus $e$ (orbital eccentricity) diagram (bottom panel of Figure \ref{Toomre_DQ}), thin disc stars occupy region A (low eccentricity), thick disc stars region B (higher eccentricity and lower $J_z$), and halo stars region C (see \cite{Pauli2006}). The location of C-enriched DQs in this plot aligns with their positions in the Toomre diagram and confirms that the kinematical ages of these WDs are consistent with those of old Galactic populations. 
Binary population synthesis calculations (\cite{KawkaNonexplosive2023}) are also in general agreement with observations and support the scenario that The carbon-rich DQs originate from the merger of two WDs, as their initial separation gradually decreases over billions of years due to GW radiation. The lower panel of Figure \ref{fig_age_obs} presents the simulated magnitude-limited population from double WD mergers with Gaia magnitude $G < 20$. Figure \ref{delay_times} shows the distribution of delay times of double WD mergers within 1\,kpc from Earth and for only one starburst. Thus, the histograms shown in Figures \ref{fig_age_obs} and \ref{delay_times} highlight that the merging of two WDs has delay times ranging from a few hundred Megayears to a Hubble time and thus they are the likely result of the mergers of two WDs (see \cite{KawkaNonexplosive2023} for all the details). 

It is a common misconception that the common envelope phase always brings stars so close together that, if they do not merge during the common envelope phase, they inevitably merge within a few hundred million years. This is not the case. A significant fraction of binaries that experience two common envelope phases evolve into double WD systems, where the separation between the two WDs gradually decreases due to GW radiation. Once the WDs are sufficiently close, a phase of mass transfer may commence, potentially culminating in a merger that could result in a supernova explosion or other outcomes (see Section \ref{SN_Progenitors}) within a Hubble time. However, there is also a significant fraction of double WD binaries that have undergone two common envelope phases but will not come into contact within a Hubble time. One example is NLTT\,12758, which contains two WDs with masses of 0.83\msun and 0.9\msun and an orbital period of 1.154 days \cite{Kawka2017_NLTT12758}.

What we have learnt from the study of C-enriched DQs, is that rather than triggering a SN\,Ia event, the merging of two WDs may result in the formation of a more massive, rapidly rotating, and often magnetic WD, thus becoming ``failed SNe\,Ia'' (\cite{Dunlap2015}). The object formed through this double WD merger may pass through an ``R Coronae Borealis'' stellar phase before ultimately evolving into a WD (\cite{Ruiter2025}). 

The space density of Carbon-rich DQs, between 0.8 and 3.0 $\times$ 10$^{-5}$\,pc$^{-3}$ (\cite{KawkaNonexplosive2023}), suggests that they could account for a significant proportion of double WD merger remnants and could provide a unique opportunity to study the physical processes that occur during the merging of WDs. Furthermore, their rapid rotation and strong magnetic fields (\cite{Dufour2013,KawkaNonexplosive2023}) tell us that merger dynamics play a critical role in generating magnetic fields through dynamo processes (\cite{Tout2008,Wick14}), thus presenting the best observational evidence that magnetic fields are generated during merging (\cite{FerrarioWickKawka_Review2020},\cite{ferrarioMagneticFieldGeneration2015}). The presence of carbon in their atmospheres, which cannot be explained by traditional convective dredge-up models, may also suggest that nuclear processing might occur during merging. Thus, the study of their birthrate and mass distribution provides a unique opportunity to constrain possible progenitor scenarios in SNe\,Ia  (\cite{Dunlap2015}).

\subsubsection{A case study: the DQ WD LP 93-21 as an ancient double degenerate merger}\label{LP93-21}

\begin{figure}
\centering
    \includegraphics[viewport=15 400 565 690,clip,width=0.75\columnwidth]{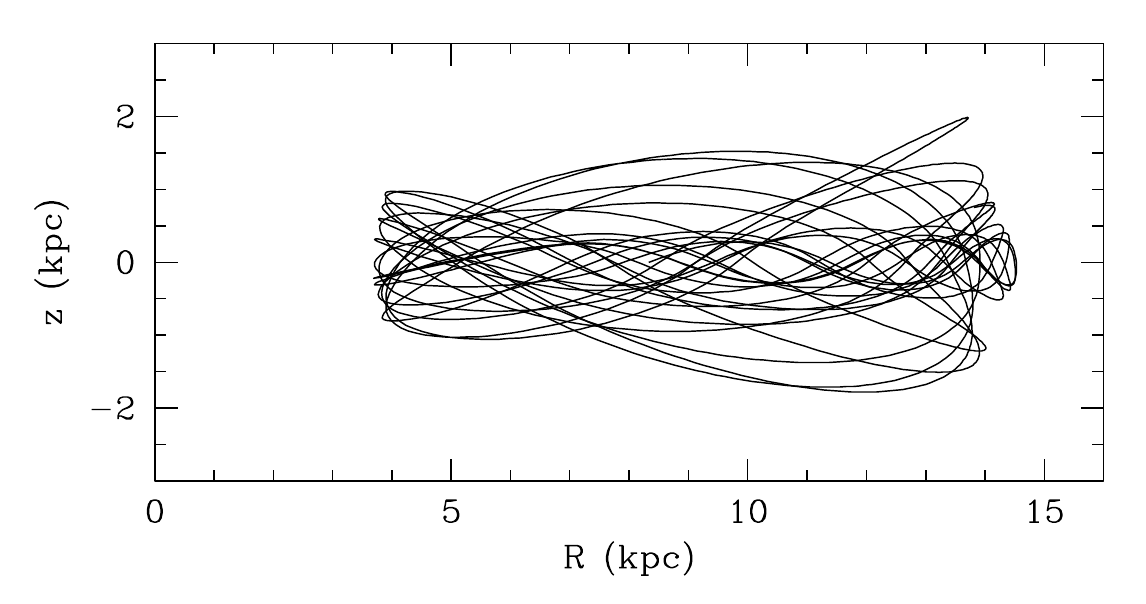}
    \caption{Orbit of LP~93-21 in the Galactic potential: Galactic height ($z$) versus radial distance $R=\sqrt{x^2+y^2}$ (from \cite{Kawka2020LP93}).}
    \label{kine_lp93}
\end{figure}

LP\,93-21, a C-enriched DQ WD, was discovered in 1968 in the Luyten Palomar survey (\cite{Luyten1968}) and proposed to be a hyper-runaway WD and a potential progeny of a Type SNIax by \cite{ruffiniHyperrunawayWhiteDwarf2019}. While LP\,93-21 is somewhat similar to the hyper-velocity WD LP\,40-365 (\cite{Vennes2017}, see section \ref{survivors}), significant differences exist (\cite{Kawka2020LP93}. The following are the characteristics of LP\,93-21 that make this WD stand out from the general population of single WDs:
\begin{enumerate}
    \item LP\,93-21 is very massive ($\sim 1.1$\,\msun) with a helium-dominated atmosphere rich in carbon \cite{Leggett2018, Kilic2019}.
    \item The effective temperature of LP\,93-21 is $\sim 9,360$\,K, and its cooling age $\sim 2.3$\,Gyr.
    \item The Galactic motion of LP\,93-21, with $(U, V, W) = (-201 \pm 11, -420 \pm 7, 57 \pm 16)$\,km\,s$^{-1}$ (\cite{Kawka2020LP93}), points to an old stellar population (\cite{Kawka2020LP93}, see Fig. \ref{kine_lp93}). The location of LP\,93-21 in the Toomre diagram and in the $J_z$ versus $e$ plot in Figure \ref{Toomre_DQ} confirms its ancient origin.
\end{enumerate}
LP\,93-21 could be the progeny of a binary system consisting of two main-sequence stars with masses of 2.1 and 1.8\,\msun in a wide orbit ($P \sim 1184$\,days). The primary evolves into an AGB star, initiating a common envelope phase, after which it becomes a CO WD. The secondary star also undergoes common envelope evolution and becomes a CO WD. Over $\sim 9.4$\,Gyr, GW emission shrinks their orbit until the two WDs merge, forming a 1.1\,$M_\odot$ WD (\cite{Kawka2020LP93}).

Interestingly, the retrograde orbit of LP\,93-21 points to a possble extra-Galactic origin, that is, this object might have belonged to a dwarf galaxy that was accreted by the Milky Way more than 10 billion years ago (redshift $z > 1.5$). 

As the likely product of an old double WD merger, LP\,93-21 exemplifies the broader population of C-enriched DQs. 

\subsection{Origin of High Field Magnetic White Dwarfs}\label{magnetic_fields}

It has been proposed that the very high fields observed in some magnetic WDs (HFMWDs) are generated by an $\alpha-\Omega$ dynamo driven by differential rotation during merging events (\cite{Wick14, Tout2008}). Population synthesis studies by \cite{Briggs2015} found that such mergers can explain the observed incidence of magnetism among WDs and their mass distribution. Furthermore, their results indicate that while the majority of HFMWDs form during a phase of common-envelope evolution, a smaller fraction arises from the merger of two WDs. They conclude that it is the merger route that gives rise to the strongest fields in WDs.

A notable example of the outcome of the double degenerate merger path is RE\,J0317-853 \cite{Barstow1995_0317, Ferrario1997_J0317, Vennes2003_J0317}. The main lines of evidence in support of its merger origin include the following. Firstly, the mass of RE\,J0317-853 is estimated to be near (or even above) 1.35\,\msun, larger than the typical upper limit for WDs formed through single star evolution. Secondly, RE\,J0317-853 rotates with a period of $\sim 725$ seconds, which is exceptionally fast for a WD. Thirdly, it possesses a magnetic field strength of $\sim 340$\,MG, placing it among the most strongly magnetic WDs known.  

Additionally, and most importantly, the age discrepancy with its less massive (non-interacting) WD companion, LB\,9802 which appears to be older than RE\,J0317-853, further supports the hypothesis that RE\,J0317-853 is the outcome of a double WD merger (\cite{Ferrario1997_J0317}).

Whilst various mechanisms have been proposed to explain the origin of magnetic fields in WDs (see, e.g., \cite{ferrarioMagneticFieldGeneration2015} for a review, and also \cite{Isern2017_Magnetism_Crystallisation} for an alternative mechanism), all of which may contribute to the population of magnetic WDs, the double-degenerate merger route remains the most promising to explain the generation of the highest field strengths (e.g., \cite{Ferrario2015_MWDs_Review}, \cite{Schmidt2001_magnetic}).

\section{The Progenitors of Supernovae of Type Ia}\label{SN_Progenitors}

It is now quite well established that SNe\,Ia result from runaway thermonuclear explosions of WDs in binaries, leading to the disruption of the WD and sometimes also that of its companion (see \cite{Ruiter2025} for a very comprehensive review). However, unlike core-collapse SNe, their progenitor systems remain poorly understood. Key uncertainties include the nature of the companion star transferring mass to the WD, i.e., whether it is a non-degenerate star (single degenerate channel) or another WD (double degenerate channel), how mass transfer occurs (thermal or dynamical timescale), the WD's mass at the time of explosion, the chemical composition of the transferred material, and the mass ratio $q=M_2/M_1$, where $M_1$ and $M_2$ are the masses of the accretor and donor, respectively. Hydrogen and/or helium are most commonly transferred in the single-degenerate channel (not considered here), whereas helium and carbon-oxygen are dominant in the double degenerate channel. Thus, understanding the contributions of all these quantities is crucial to explain the explosion mechanisms, progenitor population characteristics, incidence, and their observational signatures. 

Because SNe\,Ia exhibit a well-known relation between the width of their light curves and peak absolute magnitude (Phillips relation; \cite{phillipsAbsoluteMagnitudesType1993}), they can be used as standard candles to determine cosmological distances and to study the accelerating expansion of the universe (\cite{maozObservationalCluesProgenitors2014}). Additionally, they contribute significantly to the synthesis of rare heavy elements and provide constraints to galactic chemical evolution models (\cite{eitnerObservationalConstraintsOrigin2020}).

The most common mergers that lead to SNe\,Ia events are thought to involve two CO\,WDs (\cite{maozObservationalCluesProgenitors2014}), whilst those entailing the presence of an ONe\,WD primary are less likely to result in SNe\,Ia because, as some early studies have found (e.g., \cite{nomotoConditionsAccretioninducedCollapse1991} ), the fate of an accreting ONe\,WD depends on whether nuclear energy release occurs more rapidly than electron capture behind the deflagration wave. In other words, the very high core density of ONe\,WDs increases the likelihood of neutronisation, thus leading to an accretion-induced collapse into a neutron star rather than to a SN event.

In the double degenerate scenario, the most likely avenue to SNe\,Ia events appears to consist of a primary CO\,WD with a mass between $0.8$\,\msun\ and $1.2$\,\msun, and a secondary CO\,WD with a mass between $0.6$\,\msun\ and $1.0$\,\msun\ (e.g, see SN rates from the population synthesis calculations of \cite{ruiterRatesDelayTimes2009}). Furthermore, the work of, to mention a couple, \cite{dan2012HowMergerTwo, Pakmor2011_Mergers_EqualMass_WDs} has shown that a mass ratio $q=0.5 - 1.0$ is generally favoured in merger simulations that lead to SNe\,Ia.

Another possible scenario involves mergers between a CO\,WD and a He\,WD. In these systems, the helium accreted onto the CO\,WD can detonate, compressing the CO core and potentially triggering a secondary detonation. This is known as the ``double detonation'' mechanism, as first proposed by \cite{nomotoAccretingWhiteDwarf1982, Nomoto1982_AccretingWDs_OffCentreDetonation}.

However, not all thermonuclear SNe resulting from WD explosions are identical. Certain subclasses of SNe deviate markedly from the Phillips relation, rendering them unsuitable as standard candles. Sub-classes include highly luminous events (e.g.,super-Chandrasekhar explosions \cite{Taubenberger2013}); sub-luminous 1991bg-like SNe (e.g., \cite{Doull2011,Crocker2017_DiffuseAntiMatter_SNe91bg}), and peculiar SN\,Iax (see \cite{Jha2017} for a review). SNe\,Iax are often described as ``failed'' SNe\,Ia because they do not completely unbind the WD and leave behind a partially burnt remnant. The best example of a surviving shrapnel of a failed SN\,Ia is LP\,40-365 (\cite{Vennes2017}, see section \ref{survivors}). This diversity likely arises from multiple progenitor pathways.

The simulations by \cite{Burmester2023} and \cite{Burmester2025} have investigated the potential explosive outcomes of mergers between two WDs that were previously part of a system resembling an AM Canis Venaticorum (AM\,CVn) type binary (e.g., CR\,Boo \cite{Provencal997WholeEarth_AMCVn}). In such progenitor systems, a WD accretes helium-rich material from a low-mass donor. These binaries represent the final stage of an evolutionary process that typically involves one or two episodes of common envelope evolution, leading to extremely short orbital periods, usually ranging from 10 to 65 minutes. In such a system, the WD accretes helium-rich material from a low-mass donor, representing the end state of evolutionary processes that involve one or two episodes of common envelope evolution, resulting in very short orbital periods (typically 10 to 65 minutes). The formation of such a system is theorised to occur via three primary interaction pathways: (i) the original binary system undergoes two common envelope phases and evolves into a double WD binary, with the more massive WD accreting helium from a low-mass He\,WD (this is known as the WD channel, \cite{Paczynski1967}); (ii) the system undergoes two common envelope phases evolving into a binary where a non-degenerate helium star transfers mass to a WD (He-star channel, initially suggested by \cite{IbenTutukov1987}); and (iii) the system undergoes a single common envelope phase and becomes a Cataclysmic Variable system with H-rich material transferred from the non-degenerate star to its WD companion. Over time, the donor becomes degenerate and H-deficient  (evolved CV channel, \cite{Tutukov1985}, \cite{Podsiadlowski2003}). 

It is generally accepted that mass transfer in interacting binaries remains stable if the mass ratio $q<M_2/M_1 < 2/3$ and if accretion occurs via a disc (\cite{MarshNelemansSteeghs2004}). However, \cite{shenEveryInteractingDouble2015} have convincingly argued that even disc-accreting double WD binaries with a low $q$ can merge. This is because nova-like outbursts, triggered by initial hydrogen-rich mass transfer, create friction within the expanding shell, reducing the orbital separation and accelerating mass transfer, which might ultimately lead to merging. If merging does not occur during this phase, subsequent helium-driven nova events may drive the binary towards coalescence. Thus, this work suggests that all interacting double WD binaries may eventually merge.

\begin{figure*}
    \centering
    \includegraphics[width=0.75\linewidth ]{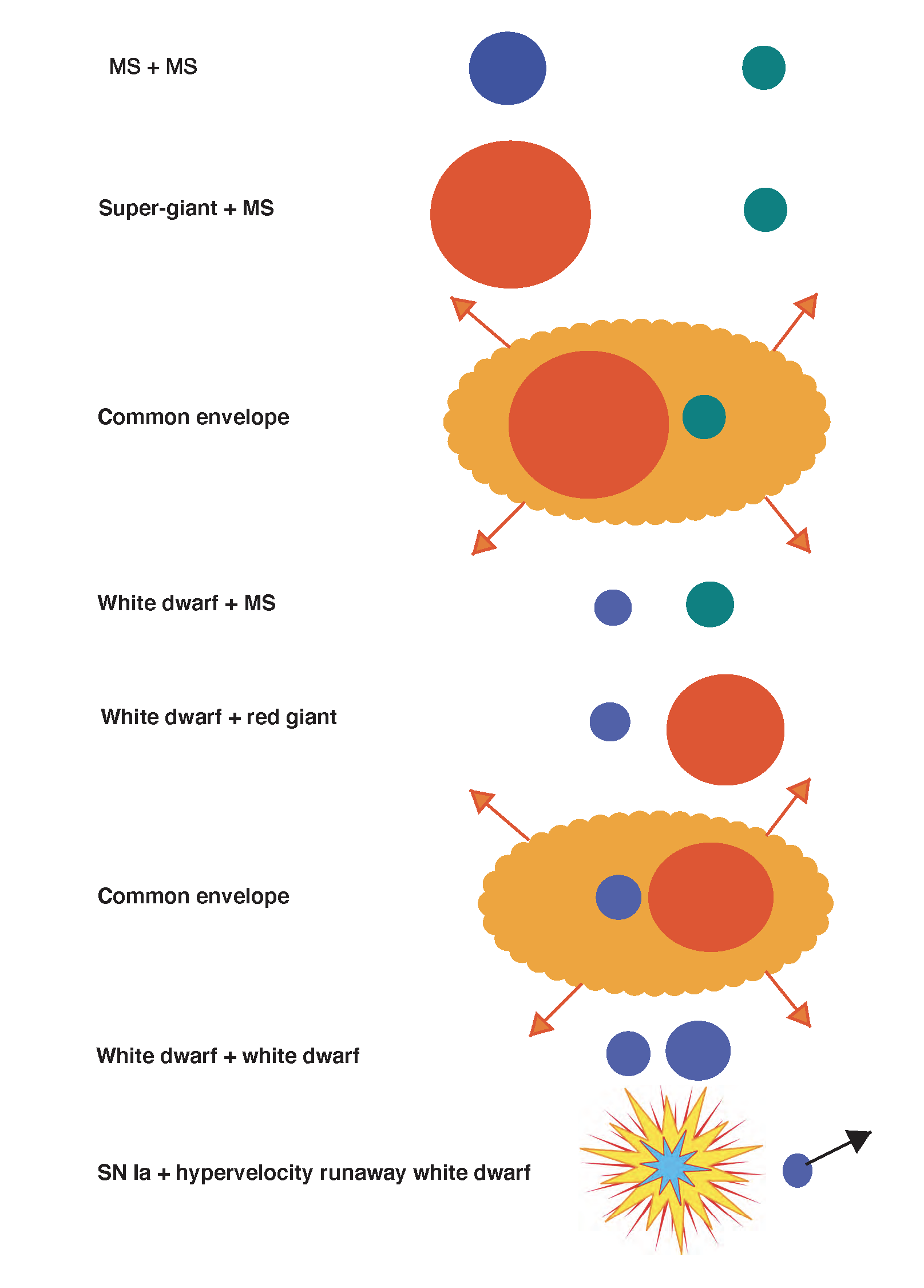}
    \caption{Evolutionary path toward an interacting AM\,CVn-like system ending with a SN\,Ia explosion and a hypervelocity runaway WD (from \cite{Burmester2023}).}
    \label{BinEvol}
\end{figure*}

The simulations of \cite{Burmester2023}, conducted with the moving-mesh code \textsc{AREPO} (\cite{springelPurSiMuove2010,pakmorImprovingConvergenceProperties2016,weinbergerAREPOPublicCode2020}), explored the possible explosive outcomes following the merger of the two WDs. In these simulations the primary is a 1.1\,\msun CO WD and its companion a 0.35\,\msun He WD. A possible progenitor system on the ZAMS could consist of a primary star with a mass in the range $5-7$\,\msun and a secondary star of $2-3$\,\msun, with orbital periods in the range $500-700$ days. The key evolutionary points are sketched in Figure \ref{BinEvol}.

The merging process results in a helium detonation at the base of the primary's helium layer that triggers an off-centre carbon detonation, leading to a SN\,Ia (see Figure \ref{BinEvol}). The expelled material mainly consists of $^{56}\mathrm{Ni}$, $^{4}\mathrm{He}$, $^{28}\mathrm{Si}$, and $^{32}\mathrm{S}$. Interestingly, these simulations predict a surviving degenerate companion with a mass of $\sim 0.23$\,$M_\odot$, travelling at velocities exceeding 1,700\,km\,s$^{-1}$, and with an atmosphere enriched with heavy elements, including approximately 0.8\% of $^{56}\mathrm{Ni}$ by mass. This finding is consistent with observations of hypervelocity WDs that have been linked to SN explosions (see \ref{survivors}). 

\begin{figure*}
\centering
\includegraphics[width=0.8\textwidth]{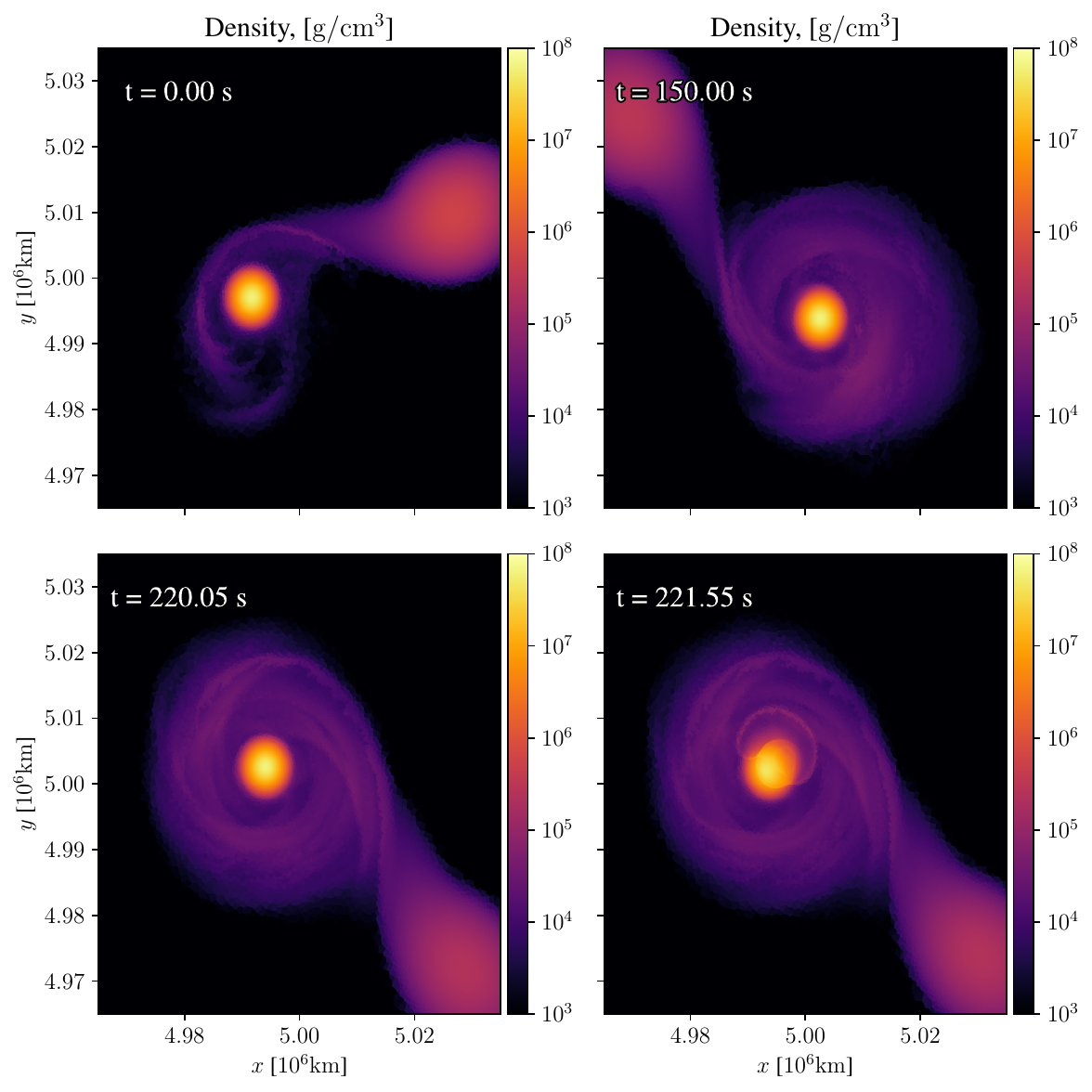}
\label{merger_simulation}
\caption{Evolution of the density during the double WD merging. Top left: Roche Lobe overflow. Top right: An accretion disc begins to form. Bottom left: The thickness of the accretion disc increases. Bottom right: A He detonation is triggered and spreads around the WD primary (from \cite{Burmester2023}).}
\end{figure*}

Similar simulations, but with the primary being a 1.1\,\msun ONe WD, were performed by \cite{Burmester2025} and showed that after the accretion of about 0.13\,\msun, helium ignites at the base of the primary's helium layer, just like in the simulations of \cite{Burmester2023}, but in this case the shell detonation cannot trigger a SN\,Ia event. The shell burning mostly produces intermediate mass elements, such as  $^{28}\mathrm{Si}$, $^{23}\mathrm{S}$ and $^{40}\mathrm{Ca}$ and a small amount of iron-peak elements (about $10^{-4}$\,\msun of $^{56}\mathrm{Ni}$). Thus, the ONe WD survives the explosion which could be classified as a sub-luminous SN event, potentially of the ``.Ia'' type. This kind of event was first envisaged by \cite{Shen2010_dotIa} in the context of helium shell detonations triggered by rapid helium accretion in  AM\,CVn systems. Observationally, SN\,2010X has been suggested as a possible candidate of a .Ia explosion (\cite{Kasliwal2010_SN2010X_dotIa}). The C-enriched DQs (see section \ref{DQ_WD}) could also be the massive WD remnants of .Ia explosions.

\subsection{Surviving White Dwarfs in Supernova Events}\label{survivors}

In many progenitor scenarios, the companion star remains intact after the WD's explosion. Efforts to identify these surviving companions have mainly focused on those associated with single-degenerate scenarios but have produced inconclusive results (e.g. \cite{PanRickerTaam2014SurvivingComanion}, \cite{Ruiz-Lapuente2018Kepler}), indicating that the single-degenerate pathway may not be the dominant route to SNe\,Ia.

Interestingly, recent discoveries show that remnants of the exploding WD itself can also survive, indicating that incomplete explosions may occur, leaving behind a fragment of the original WD (\cite{Vennes2017}).

The work of \cite{Vennes2017} has shown that LP\,40-365 is a hypervelocity object with spectra that lack hydrogen and helium lines, while exhibiting strong magnesium and sodium and weak oxygen lines. The effective temperature (10,000\,K) and gravity ($\log g=5.8$\,g\,s$^{-2}$) suggest that LP\,40-365 is a low-mass degenerate star. Its projected velocity $v\sin i = 30.5$\,km\,s$^{-1}$, where $i$ is the angle between the rotation axis and the line of sight and $v$ the equatorial rotation velocity, indicates that LP\,40-365 was spun up in a binary. This information strongly supports the hypothesis that LP\,40-365 is a stellar fragment that survived the failed or partial detonation of a SN\,Iax (\cite{Vennes2017}). Following the discovery of LP\,40-365, another three objects were identified by \cite{Raddi2019MNRAS} as belonging to the same class of partly burnt and high-velocity survivors of SN explosions. 

Another class of high-speed WD explosion survivors, known as D$^6$ stars (dynamically driven double-degenerate double-detonation stars), has been identified by \textit{Gaia} (\cite{shenThreeHypervelocityWhite2018}, \cite{El-Badry2023_D6stars}). These stars are believed to be runaway compact donors from explosive events in double WD systems.  In this scenario, the donor WD is ejected into space at its pre-explosion orbital velocity well above 1,000\,km\,s$^{-1}$. The impact of the SN ejecta onto the donor, strips away the donor's outer layers and contaminates its atmosphere with heavy elements synthesised during the explosion. Numerical simulations of merging WDs have also yielded explosions that leave the WD donor stripped and chemically polluted, but intact, and with velocities above 1,500\,km\,s$^{-1}$ (\cite{Burmester2023}, see also section \ref{SN_Progenitors}), in agreement with observations. 

Simulations of a merger between two helium-carbon-oxygen WDs (\cite{Glanz2024}) have also resulted into the partial disruption of the companion WD whose remnant core is ejected at very high speeds into space.

A comprehensive review on hypervelocity SN survivors and their evolution is given by \cite{Shen2025_Review}.

Observations of these surviving stars add valuable data to the study of SN\,Ia and sub-luminous or unusual explosive events by providing clues about the characteristics and nature of the progenitor systems. These insights are critical for improving models of SN explosions and for our  understanding of the pathways through which interacting WDs evolve into SNe\,Ia.

\section{Formation of long \texorpdfstring{$\gamma$}{gamma}-ray bursts}\label{long_bursts}

Another possible exotic outcome of double WD mergers is the formation of long $\gamma$-ray bursts (LGRBs). These objects are often associated with young star-forming regions and linked to Type Ib/c SNe, characterised by hydrogen- and helium-deficient ejecta. LGRBs are exceptionally luminous cosmological explosions, with energies (if considered isotropic) typically in the range $10^{49}-10^{54}$\,erg\,s$^{-1}$ in their prompt emission phase but with the bulk of observed events in the range of $10^{51}-10^{53}$\,erg\,s$^{-1}$ (see the review by, e.g., \cite{Levan2016LGRB_Review}).

A prominent model for formation of LGRB is the collapsar model (\cite{MacFadyen1999}) which involves the collapse of a rapidly rotating, massive star's core into a black hole, leading to the formation of relativistic jets. 

However, traditional single-star models face challenges, particularly with angular momentum loss during stellar evolution and insufficient magnetic field strength at collapse. Observations indicate that LGRBs occur more often in low-metallicity galaxies but are not exclusive to them, prompting the need for alternative formation mechanisms, such as those involving interaction and merging events in binaries. 

The binary models that are generally proposed to explain LGRBs, however, tend to focus on massive stars, which do not align with the observed displacements of LGRBs from their birth clusters. This led \cite{Tout2011} to consider the intermediate-mass binary scenario which involves stars that, as single entities, would not produce SNe but would end up as WDs. This channel provides a means to achieve rapid rotation and ultra-strong  magnetic fields (see section \ref{magnetic_fields}). Specifically, the scenario involves three common envelope stages of evolution and posits that the CO core of a naked helium star merges with an ONe WD, acquiring a strong magnetic field and subsequently collapsing into a rapidly spinning magnetar capable of launching relativistic jets. Because different amounts of carbon and oxygen can accrete onto the newly formed magnetar, runaway thermonuclear reactions generate different quantities of $^{56}$Ni, in agreement with observations of SNe associated with LGRBs. 

The formation of a rapidly spinning magnetar also raises the question of whether the currently known magnetars, displaying relatively high rotation periods of a few seconds, have spun down over time or were born slowly rotating. With an estimated birth rate of $10^{-3}$ per year, it is clear that not all magnetars can be related to LGRBs (\cite{Ghirlanda2022CosmicHistory_LGRBs}, thus suggesting there may be two types of magnetars: those spinning relatively slowly at birth and formed from single-star evolution (\cite{Ferrario2008_magnetars, Ferrario2006Fossil_Pulsars} and those resulting from the merger of two stellar cores, as proposed by \cite{Tout2011, Wick14}. The latter type of magnetars are those that should be linked to LGRBs (and potentially FRBs, see section \ref{formation_FRB}). The population synthesis models of \cite{Tout2011} predict that binaries undergoing three common envelope stages of evolution and capable of producing LGRBs are indeed rare events, in agreement with the observed rarity of LGRBs ($\approx 10^{-6}-10^{-5}$ per year per galaxy, \cite{Guetta2007ratesLGRBs, Ghirlanda2022CosmicHistory_LGRBs}). 

\section{Formation of Fast Radio Bursts}\label{formation_FRB}

Fast radio bursts (FRBs) are highly energetic radio pulses of extragalactic origin that last only a few milliseconds (\cite{lorimerBrightMillisecondRadio2007, CordesChatterjee2019, Petroff2022}), with their origins still uncertain (see \cite{Zhang2023} for a review). They are often classified into two categories: (1) repeating and (apparently) non-repeating FRBs, and (2) repeating sources, of which 50 are currently known.The repeating sources emit irregular and multiple, though unpredictable, bursts over time, with some showing frequent activities while others remaining quiescent for extended periods of time. The non-repeating sources seem to be isolated events, but it has been proposed that non-repeating FRBs may emit bursts only after extended periods of dormancy (\cite{Kirsten2024_Energy_FRBs}), thus potentially spanning durations longer than the time since such events were first identified (\cite{lorimerBrightMillisecondRadio2007}).

There appear to be notable differences between the properties of repeating and non-repeating FRBs, particularly in their dispersion measure (DM) and energy distribution. The energy distribution appears to suggest that non-repeating FRBs may represent the high-energy tail of the repeating FRB population. However, this impression could simply arise from the fact that lower-energy bursts in repeating FRBs are far more frequent than their higher-energy bursts. Consequently, one might be misled into thinking that repeating FRBs are consistently at the lower-energy end of the distribution. Nevertheless, when high-energy bursts from repeating FRBs are observed, their energy levels resemble those of non-repeating FRBs (\cite{Kirsten2024_Energy_FRBs}).

Therefore, it is essential to determine whether these observed differences arise from distinct progenitor mechanisms, viewing angles, their environment, or result from observational limitations.

While neutron stars or magnetars formed from core-collapse (CC) SNe have been commonly considered potential FRB sources (\cite{Chime2020}), another plausible formation channel involves the merger of two WDs. Such mergers, and in particular those involving massive WDs, could result in a rapid collapse and in the formation of a rapidly spinning and highly magnetic magnetar. This proposed formation mechanism is strikingly similar to the channel put forward for long gamma-ray bursts (see section \ref{long_bursts}) and thus it is possible that LGRBs and FRBs (or some sub-classes of these sources) may share a common origin. 

\cite{Kundu2020} proposed a method to determine whether repeating FRBs originate from a merger event or the core collapse (CC) of a massive star. In the merger scenario, ejected material interacts with the surrounding constant-density medium, producing shocks. Analytical modelling suggests that during the initial ``free expansion'' phase the DM and rotation measure (RM) of the radio pulses increase if the power-law index of the outer ejecta profile exceeds 6. Once the system transitions to the Sedov-Taylor phase, DM continues to increase while RM decreases, exhibiting a behaviour distinct from CC scenarios, where both DM and RM consistently decline over time. 

This variation in DM and RM evolution could be observed in known FRBs, such as FRB\,121102, which shows minor DM increases and RM decreases over time (\cite{Kundu2020}). These observations contrast with predictions from typical CC models, supporting the hypothesis that WD mergers could explain specific FRB characteristics. Thus, the medium surrounding the progenitor system plays a critical role in influencing the behaviour over time of DM and RM, underscoring the importance of studying the time evolution of repeating FRBs to identify their origins.

FRBs are associated with both young and old stellar populations, with repeating sources often associated with low-metallicity dwarf galaxies with high star-formation rates (e.g., FRB\,121102; \cite{Tendulkar2017_ApJ_FirmIdentification} and FRB\,190520B; \cite{Niu2022}). Non-repeating FRBs are more commonly found in massive, evolved galaxies with lower star-formation rates, suggesting an association with older stellar populations (\cite{Bhandari2020_NonRepeating_FRBs_HostGalaxies}). This finding, as noted by \cite{Bhandari2020_NonRepeating_FRBs_HostGalaxies}, challenges models that propose all FRBs originate from young magnetars created by super-luminous CC SNe. Instead, the diversity in the host galaxies and environments of FRBs indicates that their progenitors could be a mix of both young and old stellar populations.
\begin{figure}
\includegraphics[width=.5\linewidth]{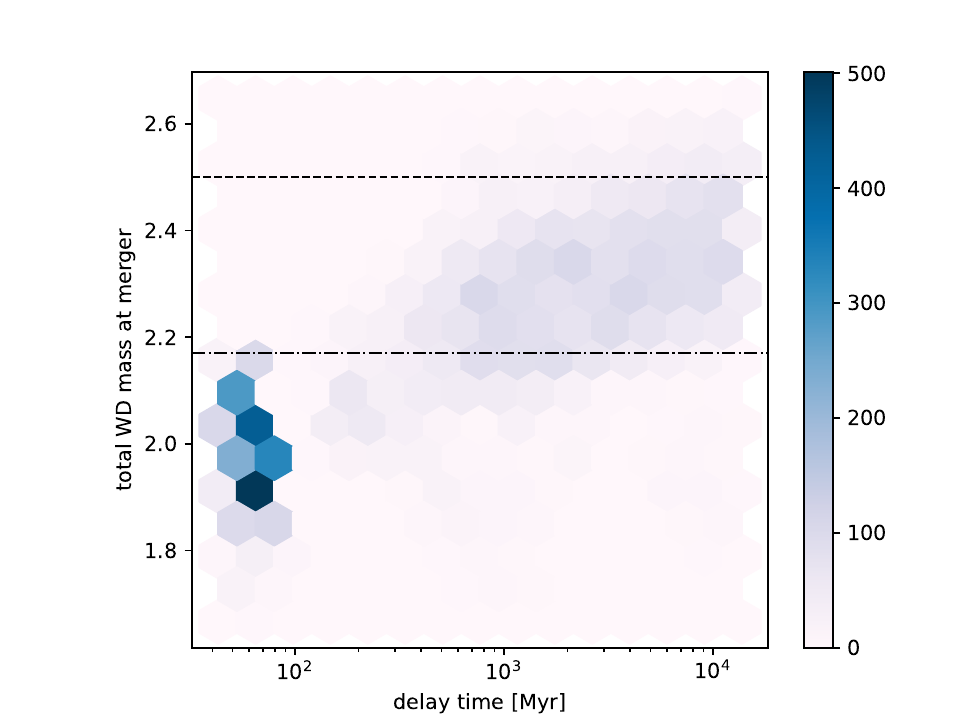}\hfill
\includegraphics[width=.5\linewidth]{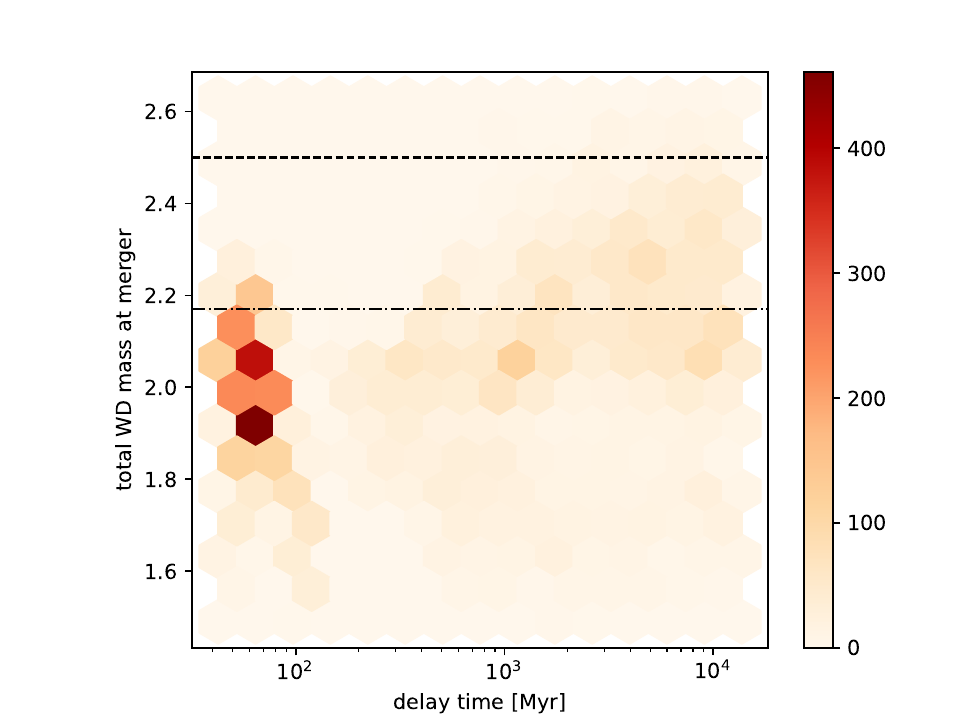}
\caption{Left: Delay times calculated using the common envelope evolution prescription described in \cite{Belczynski2008_StarTrack} with $\alpha_{\rm ce} \lambda = 1$, where $\lambda$ is the binding energy parameter and $\alpha_{\rm ce}$ the common envelope efficiency parameter (see \cite{Ruiter2019} for details). The dashed line corresponds to the neutron star upper mass limit above which the merger results into a black hole (\cite{Belczynski2008_StarTrack}). The dot-dashed line is the assumed neutron star upper mass limit as derived by \cite{margalit2017} by combining all electromagnetic and gravitational-wave data on the binary neutron star merger GW\,170817. Right: Same as left panel but with a common envelope model that takes into account the evolutionary stage of the donor star (from \cite{Ruiter2019}).}
\label{Ruiter_DelayTimes}
\end{figure}

Using population synthesis calculations, \cite{Ruiter2019} investigated binary systems that lead to neutron star formation through the collapse of ONe WDs. They examined two scenarios for such collapses: (i) accretion from a companion star, referred to as ``Accretion Induced Collapse'' (AIC), and (ii) the merger with another WD, often known as ``Merger Induced Collapse'' (MIC).

Scenario (i) is likely to result in the formation of Millisecond Radio Pulsars (MSPs), with formation pathways detailed by, e.g., \cite{Smedley2014_MSPS_HeliumCompanions, Smedley2015_redbacks}. These MSPs are expected to have magnetic fields at the lower end of the pulsar field distribution ($\lesssim {\rm a~few} \times 10^9$\,G). Although MSPs formed via AIC may be responsible for the Galactic Centre $\gamma$-ray excess signal (\cite{Gautam2022}), they are not expected to produce super-strong magnetic fields and the energy levels required to power FRBs. Scenario (ii), involving the merger of two WDs, is a plausible pathway for the formation of rapidly rotating magnetars, capable of generating the energy needed to power FRBs, and also launch the collimated jets observed in LGRBs.

The results of \cite{Ruiter2019} indicate that although ONe WDs merging with another WD tend to occur shortly after star formation, a non-negligible percentage can have delay times well exceeding the age of the Galaxy. The number density distributions of delay time against total WD mass at the time of explosion using two different approaches to account for common envelope evolution 
are shown in Figure \ref{Ruiter_DelayTimes} (\cite{Ruiter2019}). These findings strongly support the hypothesis that the formation of rapidly spinning magnetars generated through double WD mergers can occur in both young and old stellar populations, in agreement with observations of both FRBs and LGRBs.

\section{Conclusions}\label{Conclusions}

In this review I have argued that double WD mergers are responsible for many transient cosmic events.

I have summarised results indicating that the pathways proposed for SNe\,Ia can also result in the formation of HFMWDs when the conditions required for a runaway thermonuclear explosion are not met. Studying these ``failed supernovae'' provides valuable insights into the physical processes that determine the ultimate fate of merging WDs.

The DD channel for the formation of SNe\,Ia, as well as several sub-luminous events such as Type Iax,  1991bg, and .Ia explosions, is increasingly favoured over the SD channel. This shift to the DD route arises from the difficulty in detecting surviving non-degenerate companions and the observation that SNe\,Ia occur in both young star-forming galaxies and in older quiescent galaxies, as the DD pathway does not rely on ongoing star formation and thus young populations. While the SD scenario may still contribute in specific cases, the majority of SNe\,Ia are now believed to originate from the DD channel.

Sub-luminous events caused by partial burning or failed detonations during the merger process can also account for the discovery of very unusual partially burnt WD remnants.

In summary, differences in WD composition, in the combined mass at the time of explosion, mass ratio, and in the history of prior stellar interactions contribute to the diverse range of supernova luminosities and behaviours that have been observed.

Additionally, the merger of double WD systems could be an important formation channel for LGRBs and FRBs. These events could result from the rapid accretion and subsequent collapse of a rapidly rotating and highly magnetic object into a magnetar, emitting intense bursts of gamma rays and radio waves in the process. This scenario provides a plausible mechanism for generating the extreme energy outputs and fast timescales observed in these very energetic transient phenomena.

\section*{Acknowledgements}

I would like to express my gratitude to the Scientific Organising Committee of \textit{Frontier Research in Astrophysics – IV}, and in particular Franco Giovannelli, for inviting me to attend this very interesting conference and present an invited review talk on the outcomes of double degenerate mergers. Thanks also to all the organisers of this conference for their work and their hospitality in the beautiful town of Mondello in Sicily. I would also like to thank Pierre Burmester and Esha Kundu for stimulating discussions and for reading over this manuscript and providing very valuable comments and suggestions. Thanks also go to my long-term collaborators and in particular to Dayal Wickramasinghe, St\'ephane Vennes, Adela Kawka, Chris Tout, and Ashley Ruiter.

\bibliography{talk}

\bigskip
\bigskip
\noindent {\bf DISCUSSION}

\bigskip
\noindent{\bf (Q) G.\,S. Bisnovatyi-Kogan:} What numerical program was used for the simulations (seems to be 3D) of the processes in the shown movie?
\bigskip

\noindent{\bf (A) Lilia Ferrario:} The simulations were conducted by ANU PhD student Pierre Burmester using the magnetohydrodynamical moving-mesh simulation code \textsc{AREPO}, developed by Volker Springel, with significant contributions from R\"udiger Pakmor and many others.
\bigskip

\noindent{\bf (Q) Claudia Maraston:} I agree with you that the merging timescale cannot be too short as otherwise we would not have Fe-poor populations in the Universe, like elliptical galaxies.
\bigskip

\noindent{\bf (A) Lilia Ferrario:} Merging timescales can be as short as 25\,Myr to much larger than the age of the universe.
\bigskip

\noindent{\bf (Q) Silvia Zane:} Would you comment on the link between mergers and FRBs? 
\bigskip

\noindent{\bf (A) Lilia Ferrario:} WD mergers release energy on scales compatible with those required for the production of FRBs and can occur in a wide range of galaxies, from those with high star-formation rates to other with much older stellar populations. This diversity aligns with the observed distribution of FRB host galaxies. 
\bigskip

\noindent{\bf (Q) Maria Pruzhinskaya:} You have shown in the beginning that after the CE stage it takes billions of years for WDs to merge via GWs. I'd like to say that many population syntheses of binary stars (e.g., using the ``Scenario Machine'') predict that WD mergers start to occur after about 200 million years, since during the CE stage they can approach catastrophically.
\bigskip

\noindent{\bf (A) Lilia Ferrario:} While most double degenerate mergers occur shortly after star formation ($<200$,Myr), a significant fraction have delay times exceeding $10$\,Gyr. Population synthesis calculations by \cite{Ruiter2019} and \cite{KawkaNonexplosive2023}, among many others, support this conclusion. An observed double WD binary with a total mass exceeding Chandrasekhar's limit, but won't merge within a Hubble time, is NLTT\,12758 (\cite{Kawka2017_NLTT12758}).
\bigskip

\noindent{\bf (Q) S\"olen Balman:} You mentioned that magnetic fields enhance through mergers. What would you say about the enhancement of magnetic field during crystallisation of WDs. Such models are being generated by Schreiber et al. (2021-2024) and describe magnetic CVs and evolution.
\bigskip

\noindent{\bf (A) Lilia Ferrario:} Magnetic fields in WDs could be generated through various mechanisms, which may include conservation of magnetic flux from the pre-main sequence phase (fossil field hypothesis; see for instance \cite{Wick2005}), stellar mergers (\cite{Tout2008,Wickramasinghe2014}, \textit{and} crystallisation (\cite{Isern2017_Magnetism_Crystallisation}). However, in my view, it is unlikely that crystallisation can produce very strong ($1-1000$\,MG) magnetic fields. That said, the higher incidence of magnetism among older WDs compared to younger ones (\cite{Bagnulo2022_MWDs_LocalSample}) suggests that crystallisation could account for the low magnetic fields observed in Low Field Magnetic WDs (e.g., \cite{Ferrario2015_MWDs_Review}). These WDs are more prevalent in the local, volume-limited sample, which is dominated by older WDs.

Crystallisation, however, cannot explain the high magnetic fields seen in some very young magnetic WDs, such as RE\,J0317-853 ($B \sim 350$,MG) or 1RXS\,J0823.6-2525 ($B \sim 3.5$,MG). In my opinion, binary interactions remain the most promising pathway for generating MegaGauss fields in WDs, including those in magnetic cataclysmic variables.
\bigskip

\noindent{\bf (Q) Alon Retter:} You mentioned the DQ WDs group. Is there any connection to DQ Hers?
\bigskip

\noindent{\bf (A) Lilia Ferrario:} There are no links between these two groups of stars.

\end{document}